\documentclass[11pt]{article}
\usepackage[margin=1in]{geometry}
\usepackage{amsmath,amssymb,amsthm}

\newtheorem{theorem}{Theorem}
\newtheorem{lemma}[theorem]{Lemma}
\newtheorem{definition}[theorem]{Definition}

\newtheorem{remark}[theorem]{Remark}

\newcommand{\Tr}{\mathrm{Tr}}
\newcommand{\eps}{\varepsilon}
\newcommand{\cZ}{\mathcal{Z}}
\newcommand{\cX}{\mathcal{X}}
\newcommand{\vcX}{\vec{\mathcal{X}}}
\newcommand{\vcZ}{\vec{\mathcal{Z}}}
\newcommand{\uD}{\underline{D}}
\newcommand{\uI}{\underline{I}}
\newcommand{\vp}{\vec{p}}
\newcommand{\vW}{\vec{W}}
\newcommand{\Cl}{\mathrm{Cl}}
\newcommand{\cR}{\mathcal{R}}
\newcommand{\cQ}{\mathcal{Q}}
\newcommand{\vsig}{\vec{\sigma}}
\newcommand{\bX}{\boldsymbol{X}}
\newcommand{\bY}{\boldsymbol{Y}}
\newcommand{\bW}{\boldsymbol{W}}
\newcommand{\Rp}{\mathbb{R}_+^2}
\newcommand{\ket}[1]{|#1\rangle}
\newcommand{\bra}[1]{\langle#1|}

\title{Information-Spectrum General Formulas for the Capacity Region of
Classical-Quantum Multiple Access Channels}

\author{
Takuya Kubo\thanks{Independent Researcher, Yokohama, Japan. Email: \texttt{takuyakubo2010@gmail.com}}
\and
Hiroshi Nagaoka\thanks{The University of Electro-Communications, 1-5-1 Chofugaoka, Chofu-shi, Tokyo 182-8585,
Japan. Email: \texttt{nagaoka@is.uec.ac.jp}}
}
\date{}

\begin{document}
\maketitle

\begin{abstract}
For classical-quantum multiple access channels (cq-MACs) with two
senders, we give several general expressions for the capacity region,
valid for arbitrary sequences of channels. In the case of
asymptotically vanishing error ($\eps=0$), the capacity region is
expressed in terms of spectral inf-mutual information rates. For every
$0\le\eps<1$, the $\eps$-capacity region is expressed in terms of a
quantity defined through a linear combination of three alternative
hypotheses. In particular, the outer bound of Kubo and Nagaoka (2015)
is tight. Our results include the quantum counterpart of the general
formula of Han (1998) for the classical MAC. The most important ingredient is the one-shot
inner bound of Sen (2021), which we apply to each member of an
arbitrary channel sequence.
\end{abstract}


\section{Introduction}
\label{sec:intro}

The information-spectrum method, introduced by Han and Verd\'u
\cite{HanVerdu93}, is a framework that characterizes channel capacity
for arbitrary sequences of channels, without assuming stationarity or
ergodicity. For single-user channels, the general formula for the
capacity was established by Verd\'u and Han \cite{VerduHan94} in the
classical case and by Hayashi and Nagaoka \cite{HayashiNagaoka03} in
the classical-quantum case. Han \cite{Han98} obtained the general
formula for the capacity region of the classical multiple access
channel (MAC); see also \cite[Chapter~7]{HanBook}.

For the classical-quantum multiple access channel (cq-MAC), Kubo and
Nagaoka \cite{KuboNagaoka15} proved an outer bound on the capacity
region, but no matching inner bound was known. In
this paper we give several expressions for the capacity region of the
cq-MAC (Theorems~\ref{thm:main} and~\ref{thm:eps} in
Section~\ref{sec:main}). In the case of vanishing error ($\eps=0$),
the capacity region is expressed in terms of the three spectral inf-mutual information rates
(Theorem~\ref{thm:main}). This expression is the quantum counterpart of
Han's formula. For every $0\le\eps<1$, the $\eps$-capacity region is
expressed in terms of the linear-combination quantity $K$ underlying
the outer bound of \cite{KuboNagaoka15} (Theorem~\ref{thm:eps}). In
particular, that outer bound is tight. The relation to Han's results is
discussed in Remarks~\ref{rem:han} and~\ref{rem:open}.

The expression in terms of the spectral inf-mutual information rates
requires two ingredients. The most important one is the one-shot inner
bound for simultaneous decoding of a cq-MAC proved by Sen \cite{Sen21}
(Section~\ref{sec:sen}). Applying it to each member $W^{(n)}$ of an
arbitrary channel sequence yields the direct part
(Section~\ref{sec:direct}). The other is an argument that extracts the
three single-hypothesis conditions required for the direct part from
the linear-combination condition appearing in the outer bound
(Section~\ref{sec:sep}). The expression in terms of $K$ is obtained by
using the threshold test underlying the outer bound directly as a
decoder (Section~\ref{sec:qdirect}).

\section{Setup}
\label{sec:setup}

\begin{definition}[channel sequence, codes and capacity region]
\label{def:code}
Given sequences $\vcX_1=\{\cX_1^{(n)}\}_n$ and
$\vcX_2=\{\cX_2^{(n)}\}_n$ of finite sets and a sequence
$\vcZ=\{\cZ^{(n)}\}_n$ of finite-dimensional Hilbert spaces, a
\emph{cq-MAC sequence} is a sequence
$\vW=\{W^{(n)}\}_{n=1}^\infty$ of maps
\[
W^{(n)}:\cX_1^{(n)}\times\cX_2^{(n)}\to\mathcal S(\cZ^{(n)}),
\]
where $\mathcal S(\cZ^{(n)})$ denotes the set of density operators on
$\cZ^{(n)}$. No relation is assumed among the $W^{(n)}$ for different
values of $n$.

An $(n,M_1,M_2,\eps)$-code is a triple consisting of stochastic
encoders
\[
f_1(x_1\mid m_1),\ f_2(x_2\mid m_2)
\qquad(m_i\in\{1,\ldots,M_i\}),
\]
which specify the conditional probability of the input
$x_i\in\cX_i^{(n)}$ given the message $m_i$, and a decoding POVM
$\{Y_{m_1,m_2}\}$ on $\cZ^{(n)}$, such that the average error
probability
\[
\mathrm{Pe} := 1-\frac1{M_1M_2}\sum_{m_1,m_2}\sum_{x_1,x_2}
f_1(x_1\mid m_1)f_2(x_2\mid m_2)\,
\Tr[W^{(n)}_{x_1,x_2}Y_{m_1,m_2}]
\]
is less than or equal to $\eps$. For an individual cq-MAC $W$ we use
the same terminology with $n=1$. The \emph{$\eps$-capacity region}
$C(\eps\,|\,\vW)$ is the set of all $(R_1,R_2)\in\Rp$ for which there
exist $(n,M_1^{(n)},M_2^{(n)},\eps_n)$-codes, $n=1,2,\ldots$, such that
\[
\limsup_{n\to\infty}\eps_n\le\eps,\qquad
\liminf_{n\to\infty}\tfrac1n\log M_i^{(n)}\ge R_i\quad(i=1,2).
\]
Here $\mathbb R_+:=[0,\infty)$. For $\eps=0$ we simply speak of the
capacity region.
\end{definition}

\begin{definition}[marginalized states and the region $\cR$]
\label{def:region}
For sequences $\vp_1=\{p_1^{(n)}\}_n$ and $\vp_2=\{p_2^{(n)}\}_n$ of
input distributions on $\vcX_1$ and $\vcX_2$, respectively, define the
three marginalized states obtained by averaging $W^{(n)}_{x_1,x_2}$
over the input of the first sender, of the second sender, and of both
senders:
\begin{align}
\bar W^{(n)}_{p_1,x_2} &:= \sum_{x_1}p_1^{(n)}(x_1)\,W^{(n)}_{x_1,x_2},
\label{eq:tau1}\\
\bar W^{(n)}_{x_1,p_2} &:= \sum_{x_2}p_2^{(n)}(x_2)\,W^{(n)}_{x_1,x_2},
\label{eq:tau2}\\
\bar W^{(n)}_{p_1,p_2} &:= \sum_{x_1,x_2}p_1^{(n)}(x_1)p_2^{(n)}(x_2)\,
W^{(n)}_{x_1,x_2}.
\label{eq:tau0}
\end{align}
For Hermitian operators $A,B$ with spectral decomposition
$B-A=\sum_i\mu_iF_i$, we write $\{A\le B\}:=\sum_{i:\mu_i\ge0}F_i$ and
$\{A>B\}:=I-\{A\le B\}$. Using the states
\eqref{eq:tau1}--\eqref{eq:tau0}, we set
\begin{align*}
\iota_{n,1}(R\mid\vp_1,\vp_2) &:= \sum_{x_1,x_2}p_1^{(n)}(x_1)p_2^{(n)}(x_2)\,
\Tr\bigl[W^{(n)}_{x_1,x_2}\{W^{(n)}_{x_1,x_2}\le
e^{nR}\bar W^{(n)}_{p_1,x_2}\}\bigr],\\
\iota_{n,2}(R\mid\vp_1,\vp_2) &:= \sum_{x_1,x_2}p_1^{(n)}(x_1)p_2^{(n)}(x_2)\,
\Tr\bigl[W^{(n)}_{x_1,x_2}\{W^{(n)}_{x_1,x_2}\le
e^{nR}\bar W^{(n)}_{x_1,p_2}\}\bigr],\\
\iota_{n,0}(R\mid\vp_1,\vp_2) &:= \sum_{x_1,x_2}p_1^{(n)}(x_1)p_2^{(n)}(x_2)\,
\Tr\bigl[W^{(n)}_{x_1,x_2}\{W^{(n)}_{x_1,x_2}\le
e^{nR}\bar W^{(n)}_{p_1,p_2}\}\bigr],
\end{align*}
and define the three spectral inf-mutual information rates
\begin{equation}
\uI_i(\vp_1,\vp_2) := \sup\bigl\{R \mid
\lim_{n\to\infty}\iota_{n,i}(R\mid\vp_1,\vp_2)=0\bigr\}
\qquad(i=1,2,0).
\label{eq:uI}
\end{equation}
When $\vp_1,\vp_2$ are clear from the context, we omit the arguments and
write $\uI_1,\uI_2,\uI_0$. These are the quantum counterparts of
$\uI(\bX_1;\bY|\bX_2)$, $\uI(\bX_2;\bY|\bX_1)$ and
$\uI(\bX_1\bX_2;\bY)$ of Han \cite{Han98}. We define
\[
\cR(\vp_1,\vp_2) := \bigl\{(R_1,R_2)\in\Rp \;\big|\;
R_1\le\uI_1,\ R_2\le\uI_2,\ R_1+R_2\le\uI_0\bigr\}.
\]
\end{definition}

Finally, we introduce the quantities needed to state the main results
in Section~\ref{sec:main}.

\begin{definition}[the linear-combination quantity and the region $\cQ_\eps$]
\label{def:adversarial}
Let $\vp_1,\vp_2$ be sequences of input distributions on $\vcX_1$ and
$\vcX_2$, respectively. Given a sequence of triples
\[
\vsig=\{(\sigma^{(n)},\{\sigma^{(n)}_{x_1}\}_{x_1},
\{\sigma^{(n)}_{x_2}\}_{x_2})\}_n
\]
of (families of) states in $\mathcal S(\cZ^{(n)})$ and
$(R_1,R_2)\in\mathbb R^2$, we set
\begin{equation}
\Sigma^{(n)}_{x_1,x_2} :=
e^{nR_1}\sigma_{x_2}^{(n)}+e^{nR_2}\sigma_{x_1}^{(n)}
+e^{n(R_1+R_2)}\sigma^{(n)}
\label{eq:Sigma}
\end{equation}
and
\begin{align}
\kappa_n(R_1,R_2\mid\vp_1,\vp_2,\vsig) &:=
\sum_{x_1,x_2}p_1^{(n)}(x_1)p_2^{(n)}(x_2)\,
\Tr\bigl[W^{(n)}_{x_1,x_2}
\{W^{(n)}_{x_1,x_2}\le\Sigma^{(n)}_{x_1,x_2}\}\bigr],
\notag\\
K(R_1,R_2\mid\vp_1,\vp_2,\vsig) &:= \limsup_{n\to\infty}
\kappa_n(R_1,R_2\mid\vp_1,\vp_2,\vsig).
\label{eq:K}
\end{align}
For $0\le\eps<1$ we define
\begin{equation}
\cQ_\eps(\vp_1,\vp_2,\vsig) := \Rp\cap
\Cl\bigl\{(R_1,R_2)\in\mathbb R^2 \;\big|\;
K(R_1,R_2\mid\vp_1,\vp_2,\vsig)\le\eps\bigr\},
\label{eq:Qdef}
\end{equation}
where $\Cl(\cdot)$ denotes closure in $\mathbb R^2$. For the choice of
the marginalized states
\[
\vsig^{W}:=\{(\bar W^{(n)}_{p_1,p_2},
\{\bar W^{(n)}_{x_1,p_2}\}_{x_1},
\{\bar W^{(n)}_{p_1,x_2}\}_{x_2})\}_n ,
\]
we write the corresponding $\Sigma^{(n)}_{x_1,x_2}$, $\kappa_n$ and $K$
as $\Sigma^{(n),W}_{x_1,x_2}$, $\kappa_n^{W}$ and $K^{W}$.
\end{definition}

\begin{remark}[the linear-combination quantity]
\label{rem:name}
The states $\sigma^{(n)}_{x_2}$, $\sigma^{(n)}_{x_1}$ and $\sigma^{(n)}$
play the role of alternative hypotheses replacing the marginalized
states \eqref{eq:tau1}--\eqref{eq:tau0}, and $\Sigma^{(n)}_{x_1,x_2}$ in
\eqref{eq:Sigma} is a linear combination of them. Accordingly, we call
$K$ the \emph{linear-combination quantity}.
\end{remark}

\begin{remark}[the consistency constraint]
\label{rem:vsig-relax}
In \cite{KuboNagaoka15}, the sequence $\vsig$ was also required to
satisfy the consistency constraint
\[
\sigma^{(n)}=\sum_{x_1}p_1^{(n)}(x_1)\sigma_{x_1}^{(n)}
=\sum_{x_2}p_2^{(n)}(x_2)\sigma_{x_2}^{(n)}.
\]
This constraint is in fact unnecessary for the outer bound
\eqref{eq:kn} below. A self-contained proof without it is given in
Appendix~\ref{app:kn}.
\end{remark}

\section{Main results}
\label{sec:main}

\begin{theorem}[general formula, $\eps=0$]
\label{thm:main}
For every cq-MAC sequence $\vW$,
\begin{equation}
\begin{split}
C(0\,|\,\vW) &= \bigcup_{\vp_1,\vp_2}\cR(\vp_1,\vp_2)\\
&= \bigcup_{\vp_1,\vp_2}\bigcap_{\vsig}\cQ_0(\vp_1,\vp_2,\vsig)\\
&= \bigcup_{\vp_1,\vp_2}\cQ_0(\vp_1,\vp_2,\vsig^{W}),
\end{split}
\label{eq:main}
\end{equation}
where the unions run over sequences of input distributions and the
intersection over sequences of triples of states.
\end{theorem}

\begin{theorem}[general formula, $0\le\eps<1$]
\label{thm:eps}
For every cq-MAC sequence $\vW$ and every $0\le\eps<1$,
\[
C(\eps\,|\,\vW) = \bigcup_{\vp_1,\vp_2}\bigcap_{\vsig}
\cQ_\eps(\vp_1,\vp_2,\vsig)
= \bigcup_{\vp_1,\vp_2}\cQ_\eps(\vp_1,\vp_2,\vsig^{W}).
\]
\end{theorem}

The inclusion corresponding to the converse part,
\begin{equation}
C(\eps\,|\,\vW) \subset
\bigcup_{\vp_1,\vp_2}\bigcap_{\vsig}\cQ_\eps(\vp_1,\vp_2,\vsig)
\qquad(0\le\eps<1),
\label{eq:kn}
\end{equation}
was proved by Kubo and Nagaoka \cite{KuboNagaoka15} under the
consistency constraint of Remark~\ref{rem:vsig-relax}. A proof in the
present form is given in Appendix~\ref{app:kn}. Given \eqref{eq:kn}, it
suffices to show the following three inclusions:
\begin{gather}
\bigcup_{\vp_1,\vp_2}\cR(\vp_1,\vp_2) \subset C(0\,|\,\vW),
\label{eq:direct}\\
\cQ_0(\vp_1,\vp_2,\vsig^{W}) \subset \cR(\vp_1,\vp_2)
\quad\text{for all }\vp_1,\vp_2,
\label{eq:sep}\\
\cQ_\eps(\vp_1,\vp_2,\vsig^{W}) \subset C(\eps\,|\,\vW)
\quad\text{for all }\vp_1,\vp_2\text{ and }0\le\eps<1.
\label{eq:qdirect}
\end{gather}
The inclusion \eqref{eq:direct} is proved in Section~\ref{sec:direct},
\eqref{eq:sep} in Section~\ref{sec:sep}, and \eqref{eq:qdirect} in
Section~\ref{sec:qdirect}. Sen's one-shot inner bound, the key to
\eqref{eq:direct}, is stated in Section~\ref{sec:sen}.

\begin{remark}[relation to Han's general formula]
\label{rem:han}
For the classical MAC $\bW$, Han \cite[Theorem~1]{Han98} showed that
the capacity region is
\[
C(\bW)=\bigcup_{\bX_1,\bX_2}\cR_{\bW}(\bX_1,\bX_2),
\]
where the union runs over pairs of independent input processes
$\bX_1,\bX_2$, and $\cR_{\bW}(\bX_1,\bX_2)$ is defined by
$\uI(\bX_1;\bY|\bX_2)$, $\uI(\bX_2;\bY|\bX_1)$ and
$\uI(\bX_1\bX_2;\bY)$ in the same way as $\cR(\vp_1,\vp_2)$. The
equality
\[
C(0\,|\,\vW)=\bigcup_{\vp_1,\vp_2}\cR(\vp_1,\vp_2)
\]
in Theorem~\ref{thm:main} is its quantum counterpart.
Theorem~\ref{thm:eps} can be regarded as a quantum counterpart of
Han's general formula \cite[Theorem~5]{Han98} for the $\eps$-capacity
region in terms of the quantity $J$, whose role is played here by $K$.
\end{remark}

\begin{remark}[connecting the two expressions]
\label{rem:open}
Theorem~\ref{thm:main} expresses the capacity region $C(0\,|\,\vW)$
both in terms of the spectral inf-mutual information rates and in
terms of $K$. In the classical case, $J$ is the limit superior of the
probability of the union of three information-spectrum events, and the
passage from $J$ to the spectral inf-mutual information rates is
elementary. In the quantum case, $K$ is defined through a single
linear combination of three alternative hypotheses. The inclusion
\eqref{eq:sep} is elementary (Section~\ref{sec:sep}), but the reverse
inclusion is known only after taking the union over $\vp_1,\vp_2$, via
Sen's inner bound in \eqref{eq:direct} and the outer bound
\eqref{eq:kn}. It is open whether
$\cR(\vp_1,\vp_2)\subset\cQ_0(\vp_1,\vp_2,\vsig^{W})$ holds for each
$\vp_1,\vp_2$, and whether a more direct argument exists. For
$0<\eps<1$, only the expression in terms of $K$ is available.
\end{remark}

\section{Sen's one-shot inner bound}
\label{sec:sen}

The direct part \eqref{eq:direct} rests on the inner bound for
simultaneous decoding proved by Sen \cite{Sen21} for an individual
cq-MAC. To state it, we introduce the joint state $\rho^{X_1X_2Z}$ and
the products of its marginal states.

\begin{definition}[joint state and products of marginal states]
\label{def:DH}
For an individual cq-MAC $W$ with fixed input distributions $p_1,p_2$,
define the joint state
\[
\rho=\rho^{X_1X_2Z} := \sum_{x_1,x_2}p_1(x_1)p_2(x_2)\,
\ket{x_1}\bra{x_1}\otimes\ket{x_2}\bra{x_2}\otimes W_{x_1,x_2},
\]
where $\{\ket{x_1}\}_{x_1}$ and $\{\ket{x_2}\}_{x_2}$ are families of
orthonormal vectors, and the three products of its marginal states
$\tau_1:=\rho^{X_1}\otimes\rho^{X_2Z}$,
$\tau_2:=\rho^{X_2}\otimes\rho^{X_1Z}$,
$\tau_0:=\rho^{X_1}\otimes\rho^{X_2}\otimes\rho^Z$.
\end{definition}

\begin{theorem}[Sen \cite{Sen21}, two-sender case of Theorem 2$'$]
\label{thm:sen}
There exists a positive constant $c$ with the following property.
Let $W$ be an individual cq-MAC, let $p_1,p_2$ be input distributions, let
$0<\eps<1$, let $M_1,M_2$ be positive integers, and set
$M_0:=M_1M_2$. If there exist operators (tests) $0\le\Lambda_i\le I$
such that, for each $i\in\{1,2,0\}$,
\begin{align}
&\Tr[\Lambda_i\rho]\ge1-\eps,
\label{eq:sen1}\\
&\log M_i \le -\log\Tr[\Lambda_i\tau_i]-1-\log\tfrac1\eps,
\label{eq:sen2}
\end{align}
then there exists an $(n{=}1,M_1,M_2,c\sqrt\eps)$-code for $W$.
\end{theorem}

\begin{remark}[the constant in Sen's theorem]
\label{rem:senconst}
Sen \cite{Sen21} states Theorem~\ref{thm:sen} with $c=49$. We could
not verify the proof as written, but we have verified that the
statement holds with the weaker constant $c=240$. Since the
verification is lengthy, we do not include it here. None of our
results depends on the value of $c$, and only the existence of some
positive constant $c$ is used below.
\end{remark}

\section{Proof of \eqref{eq:direct}}
\label{sec:direct}

In this section we prove the inclusion \eqref{eq:direct} of the direct
part by applying Theorem~\ref{thm:sen} to each member $W^{(n)}$ of the
channel sequence separately. We use the following standard facts,
which follow from a diagonal argument:
\[
C(0\,|\,\vW)=\bigcap_{\eps'>0}C(\eps'\,|\,\vW),
\]
and $C(0\,|\,\vW)$ is closed.

\begin{proof}[Proof of \eqref{eq:direct}]
Fix $\vp_1,\vp_2$ and set
\[
\cR^\circ := \bigl\{(R_1,R_2) \;\big|\;
0<R_1<\uI_1,\ 0<R_2<\uI_2,\ R_1+R_2<\uI_0\bigr\}.
\]
We first show that $\Cl(\cR^\circ)\subset C(0\,|\,\vW)$.

To the $n$-th channel $W^{(n)}$, associate the states $\rho,\tau_i$ of
Definition~\ref{def:DH} under the input distributions
$p_1^{(n)},p_2^{(n)}$, and denote them by $\rho^{(n)},\tau_i^{(n)}$.
Since $\rho^{(n)}$ is block-diagonal in the $X_1,X_2$ basis, with
blocks $p_1^{(n)}(x_1)p_2^{(n)}(x_2)W^{(n)}_{x_1,x_2}$, and since the
corresponding blocks of $\tau_1^{(n)},\tau_2^{(n)},\tau_0^{(n)}$ are
$p_1^{(n)}(x_1)p_2^{(n)}(x_2)\bar W^{(n)}_{p_1,x_2}$,
$p_1^{(n)}(x_1)p_2^{(n)}(x_2)\bar W^{(n)}_{x_1,p_2}$,
$p_1^{(n)}(x_1)p_2^{(n)}(x_2)\bar W^{(n)}_{p_1,p_2}$, the spectral projections decompose blockwise and
\[
\iota_{n,i}(R\mid\vp_1,\vp_2)
=\Tr\bigl[\rho^{(n)}\{\rho^{(n)}\le e^{nR}\tau_i^{(n)}\}\bigr].
\]
Therefore $\uI_i$ in \eqref{eq:uI} is precisely the spectral
inf-divergence rate $\uD(\vec\rho\,\|\,\vec\tau_i)$ of Nagaoka and
Hayashi \cite{NagaokaHayashi07} for the state sequences
$\vec\rho:=\{\rho^{(n)}\}_n$ and $\vec\tau_i:=\{\tau_i^{(n)}\}_n$, and
\cite[Theorem~1]{NagaokaHayashi07} gives
\begin{equation}
\uI_i=\sup_{\{\Lambda_i^{(n)}\}_n}\Bigl\{\liminf_{n\to\infty}
-\tfrac1n\log\Tr[\Lambda_i^{(n)}\tau_i^{(n)}]\Bigm|
0\le\Lambda_i^{(n)}\le I,\
\lim_{n\to\infty}\Tr[\Lambda_i^{(n)}\rho^{(n)}]=1\Bigr\}.
\label{eq:NHtest}
\end{equation}

Let $(R_1,R_2)\in\cR^\circ$ and set $R_0:=R_1+R_2$. Then there is
$\delta>0$
with $R_i<\uI_i-\delta$ for $i=1,2,0$. Hence, by \eqref{eq:NHtest},
there exist tests $\Lambda_i^{(n)}$ such that, for every
$\eps\in(0,1)$ and all sufficiently large $n$,
\begin{equation}
\Tr[\Lambda_i^{(n)}\rho^{(n)}]\ge1-\eps,\qquad
R_i\le-\tfrac1n\log\Tr[\Lambda_i^{(n)}\tau_i^{(n)}]-\delta
\qquad(i=1,2,0).
\label{eq:test}
\end{equation}
Take $M_i^{(n)}:=\lfloor e^{nR_i}\rfloor$ ($i=1,2$) and
$M_0^{(n)}:=M_1^{(n)}M_2^{(n)}$. Then the second inequality of
\eqref{eq:test} gives
\[
\log M_i^{(n)}\le-\log\Tr[\Lambda_i^{(n)}\tau_i^{(n)}]-n\delta
\qquad(i=1,2,0).
\]
Since $-n\delta\le-1-\log\tfrac1\eps$ for all sufficiently large $n$,
the conditions \eqref{eq:sen1} and \eqref{eq:sen2} are satisfied, and
Theorem~\ref{thm:sen} yields an
$(n,M_1^{(n)},M_2^{(n)},c\sqrt\eps)$-code. Completing the sequence with
trivial codes for the finitely many remaining $n$, and using
$\tfrac1n\log M_i^{(n)}\to R_i$, we obtain
$(R_1,R_2)\in C(c\sqrt\eps\,|\,\vW)$. Since $\eps\in(0,1)$ is
arbitrary and $C(\eps'\,|\,\vW)$ is nondecreasing in $\eps'$, we
conclude that $(R_1,R_2)\in\bigcap_{\eps'>0}C(\eps'\,|\,\vW)
=C(0\,|\,\vW)$. Thus $\cR^\circ\subset C(0\,|\,\vW)$, and since
$C(0\,|\,\vW)$ is closed, $\Cl(\cR^\circ)\subset C(0\,|\,\vW)$.

If $\cR^\circ$ is nonempty, then $\Cl(\cR^\circ)=\cR(\vp_1,\vp_2)$,
so $\cR(\vp_1,\vp_2)\subset C(0\,|\,\vW)$.

If $\cR^\circ$ is empty, then $\uI_i=0$ for some $i$.
If $\uI_0=0$, or if $\uI_1=\uI_2=0$, then
$\cR(\vp_1,\vp_2)=\{(0,0)\}$, and trivial codes give
$(0,0)\in C(0\,|\,\vW)$. In the remaining case, say
$\uI_1=0<\uI_2$, every $(R_1,R_2)\in\cR(\vp_1,\vp_2)$ satisfies
$R_1=0$ and $R_2\le\uI_2$. Since $C(0\,|\,\vW)$ is closed, we may
assume $R_2<\uI_2$.
By the definition \eqref{eq:uI} of $\uI_2$, there is
$R'\in(R_2,\uI_2]$ with $\iota_{n,2}(R'\mid\vp_1,\vp_2)\to0$. Since
$\iota_{n,2}(R'\mid\vp_1,\vp_2)$ is the average over $x_1$ of the
corresponding quantity for the single-user channel
$x_2\mapsto W^{(n)}_{x_1,x_2}$ with input distribution $p_2^{(n)}$, for
each $n$ there is $x_1^{(n)}$ for which this quantity is at most
$\iota_{n,2}(R'\mid\vp_1,\vp_2)$. Hence the spectral inf-mutual
information rate of the single-user channel sequence
$\{x_2\mapsto W^{(n)}_{x_1^{(n)},x_2}\}_n$ with input distributions
$\vp_2$ is at least $R'>R_2$, and the direct part of the single-user
general formula \cite{HayashiNagaoka03} gives $(0,R_2)\in C(0\,|\,\vW)$.

Taking the union over $\vp_1,\vp_2$ yields \eqref{eq:direct}.
\end{proof}

\section{Proof of \eqref{eq:sep}}
\label{sec:sep}

Throughout this section we fix $\vp_1,\vp_2$ and omit them from the
arguments of $\kappa_n^{W}$ and $K^{W}$, writing
$\kappa_n^{W}(R_1,R_2)$ and $K^{W}(R_1,R_2)$.

\begin{proof}[Proof of \eqref{eq:sep}]
For $(R_1,R_2)\in\mathbb R^2$, $n\ge1$ and $(x_1,x_2)$, set
\[
T^{(n)}_{x_1,x_2}:=\{W^{(n)}_{x_1,x_2}>\Sigma^{(n),W}_{x_1,x_2}\}.
\]
Then
\[
\Tr[\Sigma^{(n),W}_{x_1,x_2}T^{(n)}_{x_1,x_2}]
\le\Tr[W^{(n)}_{x_1,x_2}T^{(n)}_{x_1,x_2}]\le1,
\]
and the definition \eqref{eq:Sigma} of $\Sigma^{(n),W}_{x_1,x_2}$
clearly gives
\begin{align}
\Tr[\bar W^{(n)}_{p_1,x_2}T^{(n)}_{x_1,x_2}]&\le e^{-nR_1},
\label{eq:tb1}\\
\Tr[\bar W^{(n)}_{x_1,p_2}T^{(n)}_{x_1,x_2}]&\le e^{-nR_2},
\label{eq:tb2}\\
\Tr[\bar W^{(n)}_{p_1,p_2}T^{(n)}_{x_1,x_2}]&\le e^{-n(R_1+R_2)}.
\label{eq:tb3}
\end{align}

Since $K^{W}\ge0$, we have
$\cQ_0(\vp_1,\vp_2,\vsig^{W})=\Rp\cap\Cl\{K^{W}=0\}$, and the three
conditions $R_1\le\uI_1$, $R_2\le\uI_2$ and $R_1+R_2\le\uI_0$ define a
closed subset of $\mathbb R^2$ whose intersection with $\Rp$ is
$\cR(\vp_1,\vp_2)$. It therefore suffices to show that
$K^{W}(R_1,R_2)=0$ implies these three conditions for every
$(R_1,R_2)\in\mathbb R^2$.

Suppose that $K^{W}(R_1,R_2)=0$, that is,
\[
\kappa_n^{W}(R_1,R_2)=
\sum_{x_1,x_2}p_1^{(n)}(x_1)p_2^{(n)}(x_2)\,
\Tr\bigl[W^{(n)}_{x_1,x_2}\{W^{(n)}_{x_1,x_2}\le
\Sigma^{(n),W}_{x_1,x_2}\}\bigr]\longrightarrow 0 .
\]
Fix $\delta>0$ and set
$A:=W^{(n)}_{x_1,x_2}-e^{n(R_1-\delta)}\bar W^{(n)}_{p_1,x_2}$. Since
$\Tr[A\Pi]\le\Tr[A\{A>0\}]$ for every operator $0\le\Pi\le I$,
\begin{align*}
\Tr[AT^{(n)}_{x_1,x_2}]&\le\Tr\bigl[A\{A>0\}\bigr]
\le\Tr\bigl[W^{(n)}_{x_1,x_2}\{A>0\}\bigr]\\
&=1-\Tr\bigl[W^{(n)}_{x_1,x_2}\{W^{(n)}_{x_1,x_2}\le e^{n(R_1-\delta)}
\bar W^{(n)}_{p_1,x_2}\}\bigr].
\end{align*}
Rearranging gives
\[
\Tr\bigl[W^{(n)}_{x_1,x_2}\{W^{(n)}_{x_1,x_2}\le e^{n(R_1-\delta)}
\bar W^{(n)}_{p_1,x_2}\}\bigr]
\le\bigl(1-\Tr[T^{(n)}_{x_1,x_2}W^{(n)}_{x_1,x_2}]\bigr)
+e^{n(R_1-\delta)}\Tr[T^{(n)}_{x_1,x_2}\bar W^{(n)}_{p_1,x_2}] .
\]
Taking the expectation with respect to
$p_1^{(n)}(x_1)p_2^{(n)}(x_2)$ and using \eqref{eq:tb1}, we obtain,
for each $n$,
\[
\iota_{n,1}(R_1-\delta\mid\vp_1,\vp_2)
\ \le\ \kappa_n^{W}(R_1,R_2)+e^{-n\delta} .
\]
By hypothesis the right-hand side tends to $0$. Hence
$\lim_n\iota_{n,1}(R_1-\delta\mid\vp_1,\vp_2)=0$, and therefore
$\uI_1\ge R_1-\delta$ by \eqref{eq:uI}. Letting $\delta\downarrow0$ gives
$\uI_1\ge R_1$. The remaining two conditions follow in the same way,
with $e^{n(R_1-\delta)}$ replaced by $e^{n(R_2-\delta)}$ and
$e^{n(R_1+R_2-\delta)}$ and the alternative hypothesis by
$\bar W^{(n)}_{x_1,p_2}$ and $\bar W^{(n)}_{p_1,p_2}$, respectively,
using \eqref{eq:tb2} and \eqref{eq:tb3}.
\end{proof}

\section{Proof of \eqref{eq:qdirect}}
\label{sec:qdirect}

In this section we prove \eqref{eq:qdirect} without using Sen's inner
bound. Throughout, we fix $\vp_1,\vp_2$ and, as in
Section~\ref{sec:sep}, omit them from the arguments of $\kappa_n^{W}$
and $K^{W}$. The threshold test $T^{(n)}_{x_1,x_2}$ of the proof of
\eqref{eq:sep} is used directly to construct a simultaneous decoder.

\begin{lemma}
\label{lem:threshold}
Let $n\ge1$, $(R_1,R_2)\in\mathbb R^2$ and $\delta>0$. Let $T^{(n)}_{x_1,x_2}$
be as in the proof of \eqref{eq:sep}, with $(R_1,R_2)$ replaced by
$(R_1+\delta,R_2+\delta)$. Let
$M_i:=\max\{1,\lfloor e^{nR_i}\rfloor\}$, and generate the codewords
$x_1(m_1)$ and $x_2(m_2)$ ($m_i=1,\ldots,M_i$) independently according
to $p_1^{(n)}$ and $p_2^{(n)}$, respectively. Set
$T_{m_1,m_2}:=T^{(n)}_{x_1(m_1),x_2(m_2)}$ and
$G:=\sum_{m_1,m_2}T_{m_1,m_2}$, and decode with the square-root
measurement
\[
Y_{m_1,m_2}:=G^{-1/2}\,T_{m_1,m_2}\,G^{-1/2},
\]
where $G^{-1/2}$ denotes the generalized inverse square root of $G$,
taken on its support, and the remainder $I-\sum_{m_1,m_2}Y_{m_1,m_2}$
is added to $Y_{1,1}$ to complete the POVM. Then the expected average
error probability over the random codebook satisfies
\[
\mathbb E[\mathrm{Pe}]\le\kappa_n^{W}(R_1+\delta,R_2+\delta)
+13e^{-n\delta/2}.
\]
\end{lemma}

\begin{proof}
Since the rates are replaced by $(R_1+\delta,R_2+\delta)$, the bounds
\eqref{eq:tb1}--\eqref{eq:tb3} read
\begin{align}
\Tr[\bar W^{(n)}_{p_1,x_2}T^{(n)}_{x_1,x_2}]&\le e^{-n(R_1+\delta)},
\label{eq:tbs1}\\
\Tr[\bar W^{(n)}_{x_1,p_2}T^{(n)}_{x_1,x_2}]&\le e^{-n(R_2+\delta)},
\label{eq:tbs2}\\
\Tr[\bar W^{(n)}_{p_1,p_2}T^{(n)}_{x_1,x_2}]&\le e^{-n(R_1+R_2+2\delta)}.
\label{eq:tbs3}
\end{align}
Write $W_{m_1,m_2}:=W^{(n)}_{x_1(m_1),x_2(m_2)}$. For a fixed codebook,
the average error probability is
\[
\mathrm{Pe}=\frac1{M_1M_2}\sum_{m_1,m_2}
\Tr\bigl[W_{m_1,m_2}(I-Y_{m_1,m_2})\bigr].
\]
By the Hayashi--Nagaoka operator inequality
\cite[Lemma~2]{HayashiNagaoka03}, for every $c'\in(0,1]$,
\[
I-Y_{m_1,m_2}\le(1+c')(I-T_{m_1,m_2})
+\Bigl(2+c'+\tfrac1{c'}\Bigr)
\sum_{(\hat m_1,\hat m_2)\ne(m_1,m_2)}T_{\hat m_1,\hat m_2},
\]
and hence
\begin{equation}
\mathrm{Pe}\le\frac1{M_1M_2}\sum_{m_1,m_2}\Bigl\{
(1+c')\bigl(1-\Tr[W_{m_1,m_2}T_{m_1,m_2}]\bigr)
+\Bigl(2+c'+\tfrac1{c'}\Bigr)
\sum_{(\hat m_1,\hat m_2)\ne(m_1,m_2)}
\Tr[W_{m_1,m_2}T_{\hat m_1,\hat m_2}]\Bigr\}.
\label{eq:HNPe}
\end{equation}
We take the expectation over the codebook. Since the codewords are
generated independently, the expectation of
$1-\Tr[W_{m_1,m_2}T_{m_1,m_2}]$ is
\[
\alpha:=1-\sum_{x_1,x_2}p_1^{(n)}(x_1)p_2^{(n)}(x_2)\,
\Tr\bigl[W^{(n)}_{x_1,x_2}T^{(n)}_{x_1,x_2}\bigr]
=\kappa_n^{W}(R_1+\delta,R_2+\delta),
\]
and the expectation of
$\Tr[W_{m_1,m_2}T_{\hat m_1,\hat m_2}]$ is, according as only
$\hat m_1\ne m_1$, only $\hat m_2\ne m_2$, or both indices differ,
\begin{align*}
\beta_1&:=\sum_{\hat x_1,x_1,x_2}
p_1^{(n)}(\hat x_1)p_1^{(n)}(x_1)p_2^{(n)}(x_2)\,
\Tr\bigl[W^{(n)}_{x_1,x_2}T^{(n)}_{\hat x_1,x_2}\bigr],\\
\beta_2&:=\sum_{x_1,\hat x_2,x_2}
p_1^{(n)}(x_1)p_2^{(n)}(\hat x_2)p_2^{(n)}(x_2)\,
\Tr\bigl[W^{(n)}_{x_1,x_2}T^{(n)}_{x_1,\hat x_2}\bigr],\\
\beta_3&:=\sum_{\hat x_1,\hat x_2,x_1,x_2}
p_1^{(n)}(\hat x_1)p_2^{(n)}(\hat x_2)p_1^{(n)}(x_1)p_2^{(n)}(x_2)\,
\Tr\bigl[W^{(n)}_{x_1,x_2}T^{(n)}_{\hat x_1,\hat x_2}\bigr].
\end{align*}
There are $M_1-1$, $M_2-1$ and $(M_1-1)(M_2-1)$ such pairs
$(\hat m_1,\hat m_2)$, respectively, so \eqref{eq:HNPe} gives
\begin{equation}
\mathbb E[\mathrm{Pe}]\le(1+c')\alpha
+\Bigl(2+c'+\tfrac1{c'}\Bigr)\Bigl[(M_1-1)\beta_1
+(M_2-1)\beta_2+(M_1-1)(M_2-1)\beta_3\Bigr].
\label{eq:HNsplit}
\end{equation}
In $\beta_1$ the sum over $x_1$ can be carried out first, giving
$\sum_{x_1}p_1^{(n)}(x_1)W^{(n)}_{x_1,x_2}=\bar W^{(n)}_{p_1,x_2}$.
Hence, by \eqref{eq:tbs1},
\[
\beta_1=\sum_{\hat x_1,x_2}p_1^{(n)}(\hat x_1)p_2^{(n)}(x_2)\,
\Tr\bigl[\bar W^{(n)}_{p_1,x_2}T^{(n)}_{\hat x_1,x_2}\bigr]
\le e^{-n(R_1+\delta)}.
\]
Similarly, \eqref{eq:tbs2} and \eqref{eq:tbs3} give
$\beta_2\le e^{-n(R_2+\delta)}$ and
$\beta_3\le e^{-n(R_1+R_2+2\delta)}$. Since $M_i-1<e^{nR_i}$, it
follows that
\[
(M_1-1)\beta_1+(M_2-1)\beta_2+(M_1-1)(M_2-1)\beta_3\le3e^{-n\delta}.
\]
Choosing $c'=e^{-n\delta/2}$ in \eqref{eq:HNsplit} and using
$\alpha\le1$, we obtain
\[
\mathbb E[\mathrm{Pe}]\le\alpha+e^{-n\delta/2}
+3\bigl(2e^{-n\delta}+e^{-3n\delta/2}+e^{-n\delta/2}\bigr)
\le\kappa_n^{W}(R_1+\delta,R_2+\delta)+13e^{-n\delta/2}.
\qedhere
\]
\end{proof}

\begin{proof}[Proof of \eqref{eq:qdirect}]
Let $(R_1,R_2)\in\cQ_\eps(\vp_1,\vp_2,\vsig^{W})$ and $\delta>0$.
Then there is a point $(R_1',R_2')\in\mathbb R^2$ with
$K^{W}(R_1',R_2')\le\eps$ and $|R_i'-R_i|\le\delta$ for $i=1,2$.
Apply Lemma~\ref{lem:threshold} to the point $(R_1'-\delta,R_2'-\delta)$.
Then $\limsup_n\mathbb E[\mathrm{Pe}]\le K^{W}(R_1',R_2')\le\eps$, and
for each $n$ there exists a codebook whose error probability does not
exceed the expectation. Since $M_i=\max\{1,\lfloor e^{n(R_i'-\delta)}\rfloor\}$,
we have $\tfrac1n\log M_i\to R_i'':=\max\{R_i'-\delta,0\}$, and hence
$(R_1'',R_2'')\in C(\eps\,|\,\vW)$. Since $R_i\ge0$ and
$|R_i'-\delta-R_i|\le2\delta$, we have $|R_i''-R_i|\le2\delta$. By the
same diagonal argument as for $C(0\,|\,\vW)$, the set
$C(\eps\,|\,\vW)$ is closed, so letting $\delta\downarrow0$ gives
$(R_1,R_2)\in C(\eps\,|\,\vW)$.
\end{proof}

\section{Concluding remarks}
\label{sec:conclusion}

We have given several expressions for the capacity region of the
cq-MAC. In the case $\eps=0$, the capacity region is the union of the
regions $\cR(\vp_1,\vp_2)$ defined by the spectral inf-mutual
information rates (Theorem~\ref{thm:main}). This is the quantum
counterpart of Han's formula \cite[Theorem~1]{Han98}, and its proof
combines Sen's one-shot inner bound, applied to each member of an
arbitrary channel sequence, with an argument that separates the
linear-combination condition into the three single-hypothesis
conditions. For every $0\le\eps<1$, the $\eps$-capacity region is the
union of the regions $\cQ_\eps(\vp_1,\vp_2,\vsig^{W})$ defined by the
linear-combination quantity (Theorem~\ref{thm:eps}). This can be
regarded as a quantum counterpart of Han's general formula in terms of
$J$ \cite[Theorem~5]{Han98}, and it shows that the outer bound of
\cite{KuboNagaoka15} is tight.

For $\eps=0$, the two expressions are connected by an elementary
argument in the classical case, whereas in the quantum case the only
argument known to us goes through Sen's inner bound and the outer bound
(Remark~\ref{rem:open}). Whether a simpler argument exists is open.

\section*{Acknowledgments}

In carrying out this research, the authors received assistance from
Anthropic's AI assistant Claude in checking references and logical
details, and in drafting and editing the manuscript. The authors take
responsibility for the validity of the claims and the final content of
this paper.

\appendix

\section{The outer bound without the consistency constraint}
\label{app:kn}

In this appendix we prove the outer bound \eqref{eq:kn} without the
consistency constraint of Remark~\ref{rem:vsig-relax}. In fact we prove
it for a larger class of alternative hypotheses. Throughout, let
\[
\vsig=\{(\sigma^{(n)},\{\sigma^{(n)}_{x_1}\}_{x_1},
\{\sigma^{(n)}_{x_2}\}_{x_2})\}_n
\]
be an arbitrary sequence of triples of (families of) positive
semidefinite operators on $\cZ^{(n)}$ with
$\Tr[\sigma^{(n)}]\le1$, $\Tr[\sigma^{(n)}_{x_1}]\le1$ and
$\Tr[\sigma^{(n)}_{x_2}]\le1$, and define
$\Sigma^{(n)}_{x_1,x_2}$, $\kappa_n$ and $K$ by the same formulas
\eqref{eq:Sigma}--\eqref{eq:K}. Given a code as in
Definition~\ref{def:code}, its \emph{induced input distributions} are
\[
p_i(x_i):=\frac1{M_i}\sum_{m_i}f_i(x_i\mid m_i)\qquad(i=1,2),
\]
so that the induced joint input distribution is the product
$p_1\times p_2$.

\begin{lemma}[one-shot outer bound]
\label{lem:oneshotkn}
Let $W$ be an individual cq-MAC, let a code with message sizes $M_1,M_2$,
encoders $f_i(x_i\mid m_i)$, decoding POVM $\{Y_{m_1,m_2}\}$ and average error
probability $\mathrm{Pe}$ be given, and let $p_1,p_2$ be its induced
input distributions. Then, for arbitrary constants
$\gamma_1,\gamma_2,\gamma_0>0$ and arbitrary families of positive
semidefinite operators $\{\sigma_{x_2}\}_{x_2},\{\sigma_{x_1}\}_{x_1},
\sigma$ of trace at most $1$,
\[
\sum_{x_1,x_2}p_1(x_1)p_2(x_2)\,
\Tr\bigl[W_{x_1,x_2}\{W_{x_1,x_2}\le
\gamma_1\sigma_{x_2}+\gamma_2\sigma_{x_1}+\gamma_0\sigma\}\bigr]
\le\mathrm{Pe}+\frac{\gamma_1}{M_1}+\frac{\gamma_2}{M_2}
+\frac{\gamma_0}{M_1M_2}.
\]
\end{lemma}

\begin{proof}
Write $\Sigma_{x_1,x_2}:=\gamma_1\sigma_{x_2}+\gamma_2\sigma_{x_1}
+\gamma_0\sigma$ and $m=(m_1,m_2)$. For each $m$ and each pair
$(x_1,x_2)$, the inequality
$\Tr[\rho\{\rho\le\tau\}]\le(1-\Tr[\Pi\rho])+\Tr[\Pi\tau]$, valid
for every state $\rho$, every $\tau\ge0$ and every operator
$0\le\Pi\le I$ (see the proof of \eqref{eq:sep}), gives
\[
\Tr\bigl[W_{x_1,x_2}\{W_{x_1,x_2}\le\Sigma_{x_1,x_2}\}\bigr]
\le\bigl(1-\Tr[Y_m W_{x_1,x_2}]\bigr)
+\Tr[Y_m\Sigma_{x_1,x_2}] .
\]
Multiply both sides by $\frac1{M_1M_2}f_1(x_1\mid m_1)f_2(x_2\mid m_2)$
and sum over $m,x_1,x_2$. The left-hand side then becomes that of the
claim, by the definition of the induced input distributions. The average of the first
term on the right is $\mathrm{Pe}$ by definition. The average of the
second term splits into three parts. For the $\gamma_1$-part, summing
first over $x_1$ (which eliminates $f_1$) and using
$\sum_{m_1}Y_{m_1,m_2}\le I$ for each fixed $m_2$ together with
$\Tr[\sigma_{x_2}]\le1$, we obtain
\begin{align*}
\frac{\gamma_1}{M_1M_2}\sum_{m_2}\sum_{x_2}f_2(x_2\mid m_2)
\sum_{m_1}\Tr[Y_{m_1,m_2}\sigma_{x_2}]
&\le\frac{\gamma_1}{M_1M_2}\sum_{m_2}\sum_{x_2}f_2(x_2\mid m_2)
\Tr[\sigma_{x_2}]\\
&\le\frac{\gamma_1}{M_1M_2}\cdot M_2
=\frac{\gamma_1}{M_1}.
\end{align*}
The $\gamma_2$-part is at most $\gamma_2/M_2$ by the symmetric
argument. For the $\gamma_0$-part, using $\sum_{m}Y_{m}\le I$ directly
and $\Tr[\sigma]\le1$, we have
\[
\frac{\gamma_0}{M_1M_2}\sum_{m}\Tr[Y_m\sigma]
\le\frac{\gamma_0}{M_1M_2} .
\]
Adding the three parts gives the claim.
\end{proof}

\begin{theorem}[outer bound without the consistency constraint]
\label{thm:genkn}
Let $0\le\eps<1$, let $(R_1,R_2)\in C(\eps\,|\,\vW)$, let
$\vp_1,\vp_2$ be the induced input distributions of a code sequence
witnessing $(R_1,R_2)$, and let $\vsig$ be an arbitrary sequence as
above. Then
\[
(R_1,R_2)\ \in\
\Cl\bigl\{(R_1',R_2')\in\mathbb R^2\;\big|\;
K(R_1',R_2'\mid\vp_1,\vp_2,\vsig)\le\eps\bigr\}.
\]
In particular, \eqref{eq:kn} holds.
\end{theorem}

\begin{proof}
Fix $\delta>0$ and apply Lemma~\ref{lem:oneshotkn} to the $n$-th code
of the witnessing sequence with
\[
\gamma_1=e^{n(R_1-\delta)},\qquad
\gamma_2=e^{n(R_2-\delta)},\qquad
\gamma_0=e^{n(R_1+R_2-2\delta)} .
\]
The left-hand side is then exactly
$\kappa_n(R_1-\delta,R_2-\delta\mid\vp_1,\vp_2,\vsig)$, so
\[
\kappa_n(R_1-\delta,R_2-\delta\mid\vp_1,\vp_2,\vsig)
\ \le\ \eps_n+\frac{e^{n(R_1-\delta)}}{M_1^{(n)}}
+\frac{e^{n(R_2-\delta)}}{M_2^{(n)}}
+\frac{e^{n(R_1+R_2-2\delta)}}{M_1^{(n)}M_2^{(n)}} .
\]
Since $\liminf_n\tfrac1n\log M_i^{(n)}\ge R_i$, we have
$M_i^{(n)}\ge e^{n(R_i-\delta/2)}$ for all sufficiently large $n$, so
the last three terms are at most
$e^{-n\delta/2}+e^{-n\delta/2}+e^{-n\delta}\to0$. Taking $\limsup_n$
and using $\limsup_n\eps_n\le\eps$, we obtain
\[
K(R_1-\delta,R_2-\delta\mid\vp_1,\vp_2,\vsig)\ \le\ \eps .
\]
As $\delta\downarrow0$, the points $(R_1-\delta,R_2-\delta)$ converge to
$(R_1,R_2)$, which proves the claim. Since $(R_1,R_2)\in\Rp$, and since the
sequences $\vsig$ considered here include all sequences of triples of
states, $(R_1,R_2)$ lies in
$\bigcap_{\vsig}\cQ_\eps(\vp_1,\vp_2,\vsig)$, which is contained in the
right-hand side of \eqref{eq:kn}.
\end{proof}


\begin{thebibliography}{9}
\bibitem{HanVerdu93}
T.~S.~Han and S.~Verd\'u,
``Approximation theory of output statistics,''
\emph{IEEE Trans. Inf. Theory}, vol.~39, no.~3, pp.~752--772, 1993.
\bibitem{VerduHan94}
S.~Verd\'u and T.~S.~Han,
``A general formula for channel capacity,''
\emph{IEEE Trans. Inf. Theory}, vol.~40, no.~4, pp.~1147--1157, 1994.
\bibitem{Han98}
T.~S.~Han,
``An information-spectrum approach to capacity theorems for the
general multiple-access channel,''
\emph{IEEE Trans. Inf. Theory}, vol.~44, no.~7, pp.~2773--2795, 1998.
\bibitem{HanBook}
T.~S.~Han,
\emph{Information-Spectrum Methods in Information Theory}.
Berlin: Springer, 2003.
\bibitem{HayashiNagaoka03}
M.~Hayashi and H.~Nagaoka,
``General formulas for capacity of classical-quantum channels,''
\emph{IEEE Trans. Inf. Theory}, vol.~49, no.~7, pp.~1753--1768, 2003.
\bibitem{KuboNagaoka15}
T.~Kubo and H.~Nagaoka,
``A fundamental inequality for lower-bounding the error probability
for classical and classical-quantum multiple access channels and
its applications,''
\emph{IEICE Trans. Fundamentals}, vol.~E98-A, no.~12, pp.~2376--2383, 2015.
\bibitem{NagaokaHayashi07}
H.~Nagaoka and M.~Hayashi,
``An information-spectrum approach to classical and quantum
hypothesis testing for simple hypotheses,''
\emph{IEEE Trans. Inf. Theory}, vol.~53, no.~2, pp.~534--549, 2007.
\bibitem{Sen21}
P.~Sen,
``Unions, intersections and a one-shot quantum joint typicality
lemma,''
\emph{S\=adhan\=a}, vol.~46, art.~57, 2021; arXiv:1806.07278.
\end{thebibliography}
\end{document}